\documentclass[sigconf, authorversion, nonacm]{acmart}
\usepackage{enumitem}

\AtBeginDocument{%
  \providecommand\BibTeX{{%
    \normalfont B\kern-0.5em{\scshape i\kern-0.25em b}\kern-0.8em\TeX}}}

\setcopyright{acmlicensed}
\copyrightyear{2024}
\acmYear{2024}
\acmDOI{XXXXXXX.XXXXXXX}

\acmConference[UIST '24]{User Interface Software and Technology}{October 13--16,
  2024}{Pittsburgh, PA}

\acmISBN{978-1-4503-XXXX-X/18/06}

\begin{document}

\title[Traceable Text]{Traceable Text: Deepening Reading of AI-Generated Summaries with Phrase-Level Provenance Links}

\author{Hita Kambhamettu}
\email{hitakam@seas.upenn.edu}
\orcid{0000-0001-9620-1533}
\affiliation{%
  \institution{University of Pennsylvania}
  \city{Philadelphia}
  \state{PA}
  \country{USA}
}

\author{Jamie Flores}
\email{Jamie.Flores@pennmedicine.upenn.edu}
\affiliation{%
  \institution{University of Pennsylvania}
  \city{Philadelphia}
  \state{PA}
  \country{USA}}

\author{Andrew Head}
\email{head@seas.upenn.edu}
\orcid{0000-0002-1523-3347}
\affiliation{%
  \institution{University of Pennsylvania}
  \city{Philadelphia}
  \state{PA}
  \country{USA}}

\renewcommand{\shortauthors}{Kambhamettu et al.}


\definecolor{andrewpurple}{HTML}{A53DFF}
\definecolor{andrewlight}{HTML}{5AB1E0}
\definecolor{andreworange}{HTML}{E07400}
\definecolor{darkgreen}{HTML}{009B55}
\definecolor{darkblue}{HTML}{004d80}
\definecolor{magenta}{HTML}{99195d}

\newcommand\andrew[1]{}

\newcommand\maybe[1]{\textcolor{gray}{\emph{Maybe:} #1}}

\newcommand\important[1]{\textcolor{darkgreen}{#1}}
\newcommand\unimportant[1]{\textcolor{gray}{\sout{#1}}}
\newcommand\move[1]{\textcolor{andreworange}{#1}}
\newcommand{\change}[1]{\textcolor{andrewpurple}{#1}}
\newenvironment{changes}
{\begingroup\color{andrewpurple}}
{\endgroup}

\def\computer#1{{\small\texttt{#1}}}
\AtBeginEnvironment{quote}{\itshape}

\def\subparagraph#1{\textbf{#1.}}

\def\UrlBigBreaks{\do\/\do-\do\#}
\def\UrlBreaks{\do\/\do-\do\#}

\def\shortspace{\kern 0.1em}

\def\KaTeX{K\kern-.2em\raisebox{.2em}{\scriptsize A}\kern-.12em\TeX}

\definecolor{niceblue}{HTML}{8295ff}
\def\bigbox{\color{niceblue}\rule[.25ex]{1ex}{1ex} \hskip .1ex}
\def\smallbox{\hskip .25ex \color{gray}\rule[.5ex]{.5ex}{.5ex} \hskip .25ex \hskip .1ex}
\def\boxes#1#2{
\hskip .1ex 
\newcount\boxnum
\boxnum=0
\loop
\ifnum \boxnum<#1 \bigbox \else \smallbox \fi

\advance \boxnum by 1
\ifnum \boxnum<#2
\repeat
}

\newenvironment{inlinefigureenv}
{\setlength{\topsep}{2.5ex}\center}
{\endcenter}

\newcommand{\inlinefigure}[2][.5\textwidth]{%
\begin{inlinefigureenv}%
\includegraphics[width=#1]{#2}%
\vspace{-1.25ex}%
\end{inlinefigureenv}%
}

\begin{abstract}

As AI-generated summaries proliferate, how can we help people understand the veracity of those summaries? In this short paper, we design a simple interaction primitive, \emph{traceable text}, to support critical examination of generated summaries and the source texts they were derived from. In a traceable text, passages of a generated summary link to passages of the source text that informed them. A traceable text can be generated with a straightforward prompt chaining approach, and optionally adjusted by human authors depending on application. In a usability study, we examined the impact of traceable texts on reading and understanding patient medical records. Traceable text helped readers answer questions about the content of the source text more quickly and markedly improved correctness of answers in cases where there were hallucinations in the summaries. When asked to read a text of personal importance with traceable text, readers employed traceable text as an understanding aid and as an index into the source note.

\end{abstract}



\begin{teaserfigure}
  \includegraphics[width=\textwidth]{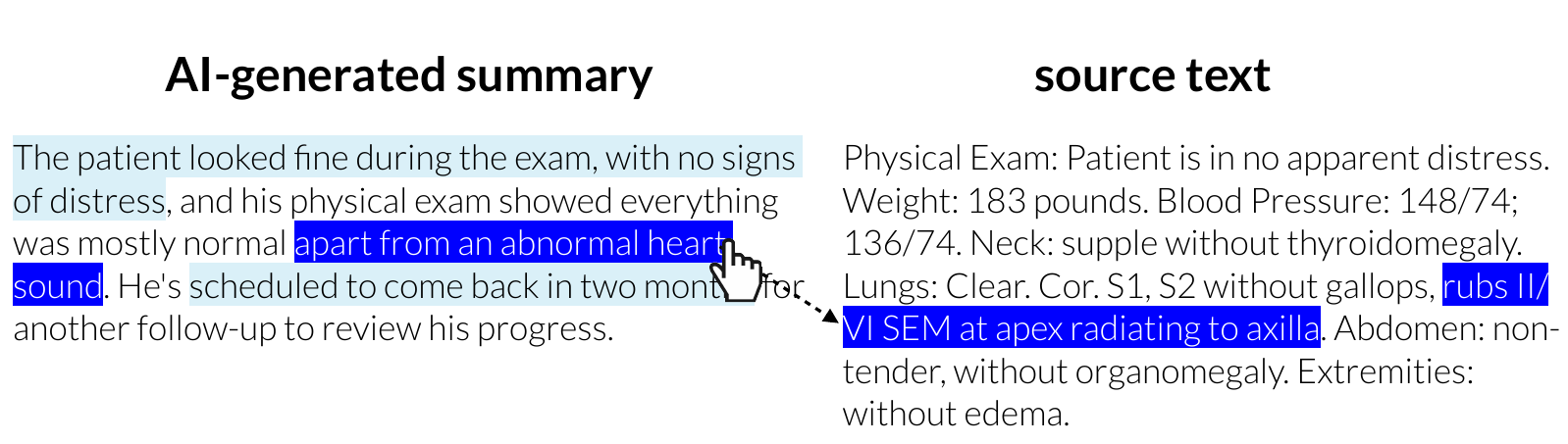}
  \caption{Interacting with a traceable text. \textmd{A traceable text is an AI-generated summary with phrase-level links to corresponding passages in a source text. Shown above is an AI-generated summary (left) of a source text (a patient medical record, right). As the reader hovers over phrases in the summary (e.g., ``apart from an abnormal heart sound''), corresponding passages from the source text are highlighted (e.g., ``rubs II/IV SEM at apex radiating to axilla'').
  The links help a reader to check the veracity of a passage in a summary, and also offer an index into the source text should the reader wish to learn more. \emph{Note}: the user interface figures in this paper are annotated reconstructions of the UI, with adjusted fonts and added dotted arrows.}}
  \vspace{4ex}
  \Description{Insert description here.}
  \label{fig:teaser}
\end{teaserfigure}

\maketitle
\pagestyle{plain}

\section{Introduction}

Advances in natural language processing (NLP) have significantly transformed the landscape of text simplification and summarization tasks, enabling AI models to distill complex information into more accessible and concise forms. From medical text~\cite{dave2023chatgpt} to legal documents~\cite{jain2021summarization}, scientific literature~\cite{semantic_scholar_tldrs_2023} to online documents~\cite{martin2023notebooklm}, interest is mounting in incorporating AI-generated summaries to help users quickly understand the essence of lengthy, complicated texts.
However, the convenience of summaries does not always eliminate the need for deeper engagement with the source material. Particularly in contexts where accuracy and detail are paramount, such as medical or legal documents, readers may need to verify the information in the AI-generated summary or explore the source text in greater depth. This necessity becomes even more critical when summaries could include hallucinatory content.

To address this challenge, this paper introduces an interaction primitive, \emph{traceable text}, designed to provide linkages between AI-generated summaries and the sources they were generated from. Traceable texts enable readers to navigate from specific claims in a summary to the corresponding passages in the source document, facilitating a critical examination of the information. This feature not only supports trust-building in AI-generated content but also empowers readers to engage more deeply with complex texts.

Traceable texts can be generated with a straightforward prompt chaining approach. The approach can be tailored to a specific domain and, as we have done ourselves, can be augmented with expert validation when appropriate. This leads to a kind of provenance links that is synthetic. While they do not necessarily maintain fidelity to how the AI generated the summary in the first place, our experience has yielded mostly-accurate generations (see Section~\ref{accuracy}).

To assess the impact of traceable text on reading experience, we conducted a usability study (see Section \ref{sec: eval}). Our study focused on a domain where source texts are complex, summaries are useful, and hallucinations matter: patients understanding their medical records~\cite{ref:kambhamettu2024explainable,esch2016engaging}. Participants were asked to read patient medical records with accompanying AI-generated summaries augmented with traceable text and answer questions that required understanding of summaries and source texts. To reflect a variety of contexts in which traceable texts might be employed, some summaries contained authentic hallucinations, and others were completely verified by experts a priori.

We found that traceable text helped readers answer questions about the content of the source text significantly more quickly. When answering questions about hallucinatory content, we observed a marked improvement in correctness (from 12.5\% to 70\%); when answering questions about verified summaries, there was a modest but statistically insignificant improvement (75\% to 90\%). Participants brought their own medical notes for a final open-ended reading task. In this task, they found traceable text useful for understanding the note and as an index into the note, and  95\% reported they could see using traceable text if it was available in a reading interface for personal medical notes.

In summary, this paper contributes the interaction primitive of the traceable text, an instantiation of its design, a description of a straightforward implementation technique, and insights from a usability study exploring its impact on reading experience.

\section{Background \& Related Work}

\subsection{Interactive summaries}

Significant research efforts have been made towards advancing automatic text summarization technologies~\cite{koh2022empirical, cohan2018discourse, dong2020discourse}. There has been considerable recent work in developing summary texts~\cite{cachola2020tldr, altmami2022automatic, tang2023evaluating} and designing interactive summaries~\cite{ref:august2023paper, fok2023qlarify} for specialized, long documents.
While summaries may be useful to readers, there are many diverse and specific informational needs of individuals that summaries might not necessarily meet. Prior HCI research has investigated engaging users in the summarization process to create summaries that better reflect individual needs~\cite{ghodratnama2020adaptive,zhang2023concepteva}. It has also created summaries that allow users to dynamically choose the level of detail they wish to see for different materials, such as lecture videos~\cite{ref:pavel2014video}, books~\cite{wu2021recursively}, films~\cite{ref:pavel2015sceneskim}, or online texts \cite{bernstein2009hypertext}. Our research extends beyond this prior work, introducing a novel interaction primitive that connects phrases in a summary with relevant passages of the source text they summarize.

\subsection{Interactive verification of generated texts}

Our work aligns with recent research that has sought to help users better verify the text generated by AI systems. Prior work has offered new mechanisms for users to validate AI-generated content, for instance by visualizing human-AI interactions involved in writing a document~\cite{ref:hoque2023hallmark}, by presenting suggested edits and evidence to users~\cite{krishna2024genaudit}, or by warning authors when an LLM introduces new information within a document being edited~\cite{laban2023beyond}. Additionally, prior research has explored how to present many LLM responses at once by helping end users gauge similarities and differences across different outputs~\cite{ilonka2024supporting}. \citet{hennigen2023towards} introduces an interaction similar to traceable text, interleaving generated text with explicit symbolic references to fields present within a data source. We explore the usability of a similar technique for phrase-level linkages. Our work complements this prior work by exploring and evaluating the particular feature of phrase-level linkages between generated summary text and the source text.

\subsection{Interlinked text and context}

Recent research has explored how to enrich the reading experience with links between a text and supporting information, whether to link texts to visualizations~\cite{ref:latif2021kori,ref:sultanum2023datatales}, data tables~\cite{ref:kim2018facilitating,ref:badam2018elastic}, code~\cite{ref:yen2023coladder}, or math formulas~\cite{ref:head2022math}. Some of this research, like ours, has observed that modern LLMs can be used to produce these linkages (e.g.,~\cite{ref:sultanum2023datatales,ref:yen2023coladder}). This prior work highlights a community interest in supporting holistic reading of complex documents via lightweight augmentations to those documents; our work shares this value and explores a related interaction primitive specifically for linking passages of an AI-generated summary to its source text.

\section{Traceable Text}
\label{sec: ttimplementation}

\subsection{Interaction design}

The purpose of traceable text is to help readers more critically examine an AI-generated summary and engage with the source text it was derived from.
More precisely, our goals are as follows:

\begin{enumerate}[label=G\arabic*.]
    \item Help readers access information in the source text.
    \item Help readers understand the source text.
    \item Help readers notice hallucinations.
    \item Employ a minimalist design.
\end{enumerate}

We designed the traceable text interaction primitive to support these goals. The design works as follows: 

\begin{figure}
    \includegraphics[width=\linewidth]{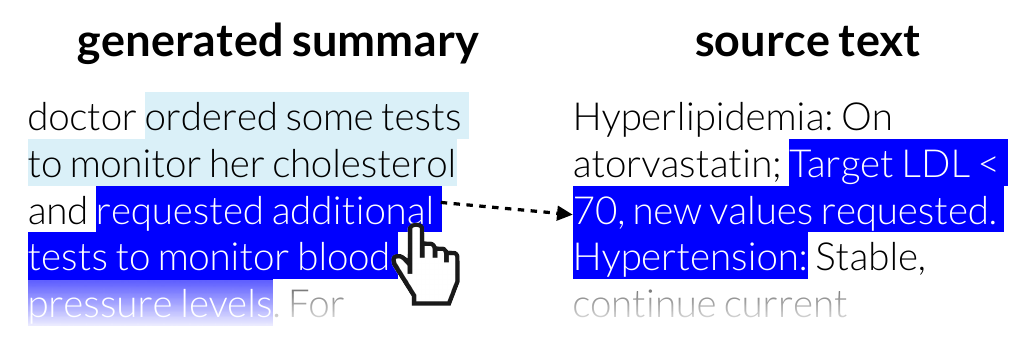}
    \caption{Inspecting hallucinations. \textmd{When a generated summary contains a hallucination (e.g., the contradiction between dark blue passages in the summary and source above), a traceable text sometimes supports resolution of the contradiction by linking between the most closely related content in the hallucinatory phrase and the original note. This can be particularly useful when contradictions are subtle to the particular reader (as in the example above if read by a non-expert patient) and might otherwise go undetected.}}
    \label{fig:hallucination}
\end{figure}

\paragraph{Integrated workspace}
The AI-generated summary and the source note are shown side by side with each other (as in Figure~\ref{fig:teaser}) so that they can be reviewed by the reader in parallel.

\paragraph{Highlighted claims}
On page load, claims in the summary are highlighted in a light blue color, similar to the color of selected text.

\paragraph{Reveal link on hover}
Whenever a user hovers over a highlighted claim, the corresponding passage that it represents from the source text is highlighted. Both the hovered claim and the source text passage are highlighted in a deeper blue color to distinguish them from the highlighted claims. This linking behavior is the mechanism by which readers can more easily access information in the source text that they think might be of interest, i.e., if they found a part of the summary they wish to know more about (G1). 

With just the design elements above come the means for noticing hallucinations (G3). A claim with a hallucination will either not be highlighted in the first place (suggesting it may have no basis in the source text), or it will connect to a related passage in the source text that may contain refuting evidence (Figure~\ref{fig:hallucination}).

\begin{figure}
    \includegraphics[width=\linewidth]{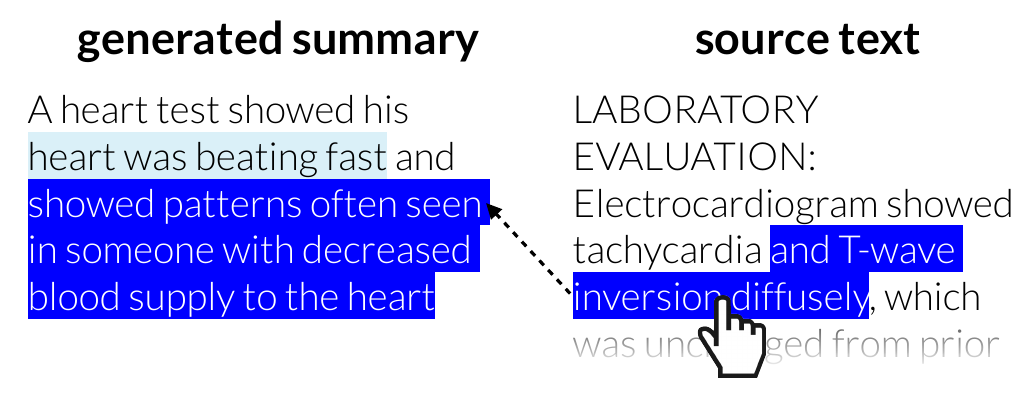}
    \caption{Backlinking from source to summary. \textmd{Readers can receive lightweight help in understanding the source text by hovering over a passage in the source text, and seeing the corresponding passage in the summary highlighted.
    }}
    \label{fig:sourcetosummary}
\end{figure}

\begin{figure}
\
    \includegraphics[width=.8\linewidth]{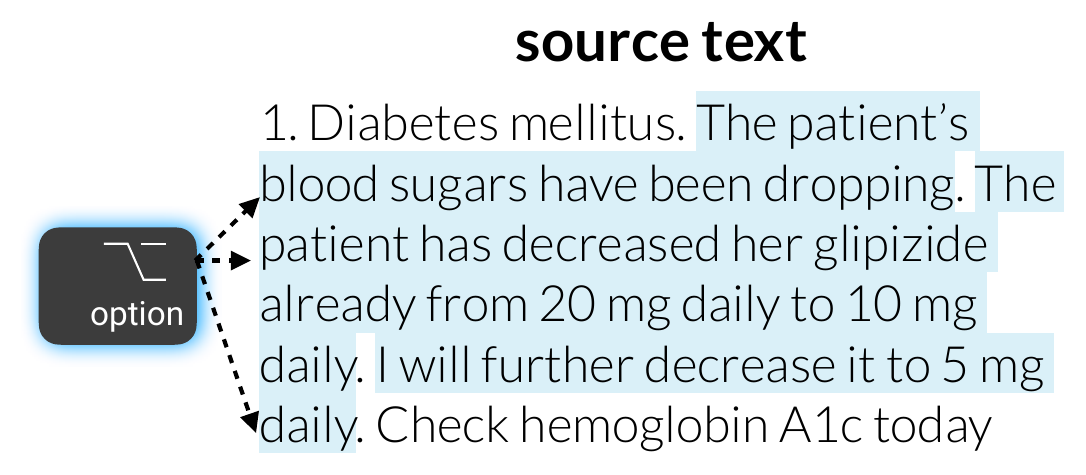}
    \caption{Showing all backlinks. \textmd{The passages of the source that are summarized are not highlighted, to avoid overwhelming the reader. If the reader wants to see passages in the source that link to the summary, they can hold a modal key, and all linked source passages become highlighted.}}
    \label{fig:backlinks}
\end{figure}

\paragraph{Backlinks}
When the reader hovers over a passage in the source text that has a corresponding claim in the summary, that claim is highlighted (Figure~\ref{fig:sourcetosummary}). In this way, readers are given more support in understanding a source text (G2), in particular for those parts that are represented in the summary. Should a reader wish to see which passages in the source text are represented in the summary, they can hold the ``Option'' key (Figure~\ref{fig:backlinks}).

\paragraph{Minimalism}
Traceable text was designed to be a lightweight and portable interaction mechanism (G4). Its only visual impact on the page is highlighting of interactive text, in two familiar colors for highlighting text. To avoid disrupting typical reading flow in the source text, backlink highlights are hidden by default.

\subsection{Implementation}

Traceable text can be implemented with a straightforward prompt chaining approach, with tailoring to domain and optional post-hoc validation, depending on application.

\begin{figure}
    \includegraphics[width=0.5 \textwidth]{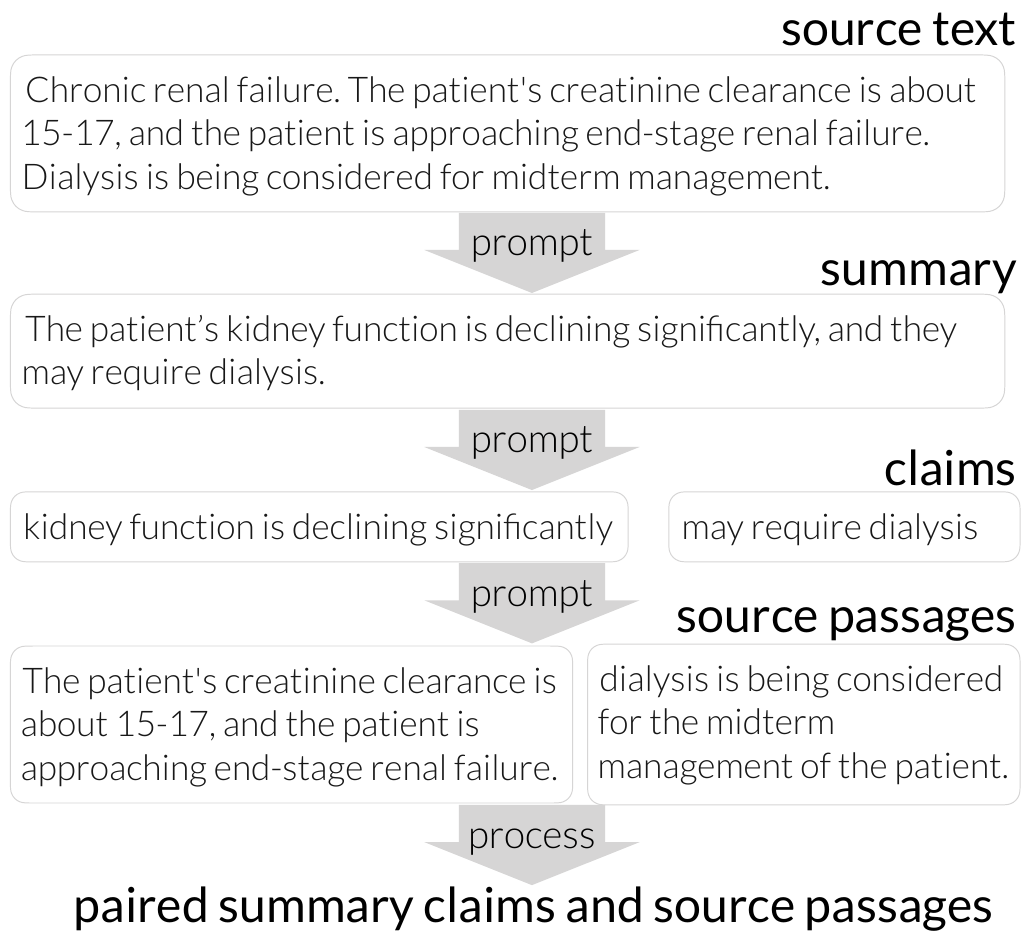}
    \caption{Prompt chain for generating traceable text. \textmd{The chain generates a summary, splits it into claims, and aligns those claims with source passages.}}
    \label{fig:implementation}
\end{figure}

\paragraph{Collecting links.} Summary-source links can be established using a chain of prompts to a large language model (LLM). In our implementation, we follow this process (see also Figure~\ref{fig:implementation}).

To summarize a source text, our implementation submits a few-shot prompt~\cite{song2023llm, brown2020language} to the OpenAI GPT-4 API~\cite{achiam2023gpt}. The prompt requests brevity and a focus on parts of the text that would be important to a non-expert reader, and discourages jargon. Then, the LLM is prompted to segment the summary into claims, and is requested to make those claims as granular as possible. The LLM is steered towards complete coverage of the summary by being given example summaries where claims cover almost all of the summary. Finally, the LLM is prompted once more to map the claims to passages in the source text. This prompt chain is implemented in a Python script.\footnote{All prompts appear in the supplemental material.}

\paragraph{UI} Our current prototype of traceable text is implemented as a React application. The system loads the source text and the AI-generated summary into side-by-side panels. Then it reads a JSON file containing outputs from the prompt chain. Summary claims and corresponding source passages are wrapped with HTML \texttt{span}s that, when hovered over, trigger the link. Because this interaction mainly requires programmatic styling of text and listening to hovering events, it could be ported to many other web-based text reading interfaces in a straightforward way.

\paragraph{Accuracy}
\label{accuracy}
How accurate is the prompt chaining approach at collecting links? 
We spot-checked accuracy on documents from our evaluation (medical progress notes). After we used the above approach to generate summaries and links, we asked clinical informatics fellows to review the accuracy of summaries and the phrase links themselves (i.e., whether they reflected the source text).

Accuracy was good, albeit not perfect. Of 159 summary-to-source links for 10 source texts, 129 were completely correct, 18 had ``semantic issues,''\footnote{``Semantic issues'' occurred when the summary claim lightly diverged from the semantics of the source passage. For example, one annotator marked a ``semantic issue'' for a claim reading `sore throat and cough' linking to `rhinopharyngitis' (a condition that often---but not always---exhibits sore throats and coughs) in the source note.} and 12 were incorrect. Incorrect links largely arose due to hallucinations in the generated summary. They also included instances where a claim linked to only part of the relevant content in the source (e.g., only 1 of 2 relevant source sentences).

\paragraph{Considering correction workflows}
In domains with sensitive content (e.g., from medicine, law), traceable texts may need to be completely free of incorrect claims or links. We piloted a correction workflow wherein we asked domain experts (clinical informatics fellows) to validate a set of generated summaries and linkages (presented to them as commented Microsoft Word documents), and then propose fixes (see also Section~\ref{stimuli}). A typical summary contained 150 words, and a typical source text contained 300 words. Validation of source/summary pairs typically took experts 10--25 minutes per pair. Our experiences suggest that experts can validate traceable text, though accelerated workflows are likely needed to validate large amounts of it (see Section~\ref{future:links}).

\section{Evaluation}
\label{sec: eval}

To evaluate the effect of traceable text on reading experience, we conducted a usability study. We had the following questions:

\medskip

\textit{RQ1: How does traceable text influence how readers notice hallucinations?} Our hypothesis was that participants would more consistently identify hallucinations in generated summaries when it was linked to related passages in the source text.

\textit{RQ2: How do phrase links affect readers' understanding of the texts?} Does traceable text help readers understand passages in the summary and source note?

\textit{RQ3: How do readers use traceable text?} When is it used during reading? What is it useful for?

\medskip

The document at the center of our study were medical progress notes---that is, notes that clinicians take during visits with patients. We choose this document as it was relatable to study participants---it is a requirement in some countries that patients can access and read their notes to review details of their care~\cite{delbanco2010open}. Furthermore, progress notes are a document where hallucinations matter. Finally, notes represented a kind of document where we could have participants bring documents of personal interest to them (i.e., their own notes).

\subsection{Methods}

\paragraph{Participants}

21 participants were recruited. Participants were recruited from several channels. Given the focus on medical documents, 16 patients were recruited from a university-affiliated patient and family advisory council. The remaining participants were recruited from academic mailing lists and social media posts.

19 of 21 participants provided demographic information. Of these, 70\% identified as female and the rest (30\%) as male. Ages ranged between 23 and 90 years old, with a median age of 60. When asked to report their occupation, 7 participants described themselves as retired, 4 as graduate students, 2 as educators, 1 as self-employed, 1 as a finance manager, 1 as an audio engineer, and 2 reported that they were disabled (1 participant did not report their occupation). Participants were compensated \$35 USD. One participant did not complete the study, and was excluded from quantitative analyses.

\paragraph{Procedure}

Our procedure received ethics review from our institution's IRB. To permit us to recruit participants who might find it prohibitive to come into the lab (namely, non-academic participants), study sessions were conducted over Zoom.

Participants first provided consent and then were led through a tutorial of traceable text's features. The study then followed a within-subjects design, and consisted of three phases.

\paragraph{\textit{Phase 1: Summaries with hallucinations.}}
Participants were asked questions about source texts and accompanying summaries where the summary contained hallucinations. For half of the questions, participants had access to traceable text; participants were counterbalanced so that each task was performed with each interface by an equal number of participants. Questions were designed to be tricky, requiring the reader to notice and resolve disagreement between the summary and  source text (more details about the hallucinations appear in Section~\ref{stimuli}). Participants answered four questions total, each about a different source/summary pair.
  
\paragraph{\textit{Phase 2: Validated summaries and links.}} 
This phase was the same as Phase 2 (involving four new but similar questions). Questions focused only on fully-verified summaries and (when using traceable texts) links. The purpose of this phase was to understand the effect on reading when traceable texts have no issues. Participants were told that all summaries had been expert-validated.

\paragraph{\textit{Phase 3: Unstructured reading time}}
Participants were given the remainder of the time (typically 5--10 minutes) to read one of their own medical notes using traceable text.\footnote{We collected and processed participants' notes prior to the study session. All participants were made aware that their notes (with identifying and sensitive information manually removed) would be passed into a commercial foundation model, and explicitly consented to process their notes.} This was intended to better reflect realistic contexts of use than Phases 1 and 2. Participants were encouraged to think aloud.

\paragraph{Questions} Questions were selected to reflect those readers may have of their medical notes, including questions about care recommendations~\cite{ref:kambhamettu2024explainable}. Answers were multi-select with 5 options each (i.e., all correct options needed to be selected). This was to incentivize readers to more closely review the summary and source text, rather than stopping after they found one answer in the text. With this design, participants had a $1/2^5$ probability of randomly guessing the right answer, or about 3\%. Questions and answers were validated by medical experts (clinical informatics fellows).\footnote{All questions and stimuli appear in the supplemental material.}

\paragraph{Measures}
For Phase 1 and 2, we measured the following:

\begin{itemize}
  \item \textit{Correctness ---} A binary variable indicating whether the participant's response to the question was correct.
  \item \textit{Speed ---} The number of seconds taken by the participant to answer the question, from when the participant first saw the question to when an answer was submitted.
  \item \textit{Difficulty ---} A five-point Likert scale variable indicating the participant's self-assessment of the following prompt: "I found this task difficult."
  \item \textit{Confidence ---} A five-point Likert scale variable indicating the participant's self-assessment of the following prompt: "I feel confident in my answer."
\end{itemize}

After completing all phases, the participant filled out a questionnaire where they reflected on the usability of traceable text.

\paragraph{Stimuli}
\label{stimuli}

Source texts were medical notes sampled from the n2c2 NLP Research dataset~\cite{kumar2015creation}, specifically the 2014 corpus of patients managing coronary artery disease diagnoses. 

For Phase 1, summaries were selected that contained real hallucinations from GPT-4. Two kinds of hallucinations were represented, because they were the hallucinations that GPT-4 produced.\footnote{To characterize the kinds of hallucinations in GPT-4 produces in summaries in this domain, we generated 34 summaries of medical notes, and asked experts (medical residents) to review them. They reported that 3 of the summaries contained fabrications and 3 included extrapolations.} The first kind of hallucination were contradictions between the summary and the source text, and the second kind were extrapolations (i.e., claims that were not necessarily incorrect, though not explicitly entailed by the source text). These represent both intrinsic and extrinsic hallucinations as they are termed in a recent review of hallucinations~\cite{ji2023survey}.

Phase 2 required expert-validated summaries and links. We recruited two medical experts---clinical informatics fellows, who are also board-certified primary care physicians---to perform this validation. Experts were paid \$60 USD/hour.

\paragraph{Analysis}

We compared correctness with and without traceable text with the $\chi^2$ test, and compared difficulty and confidence with the Wilcoxon unpaired two-sample test. Differences in time were analyzed by fitting mixed effects models, with reading interface, task number, and the interaction between them as fixed effects, and participant ID as a random effect. Statistical significance was assessed using $F$-tests with Holm-Bonferroni correction~\cite{holm1979simple} and an $\alpha$ level of .05. RQ3 was answered following a single-author thematic analysis of notes taken during study sessions. In reporting results, we refer to participants with the pseudonyms P1-P21.

\section{Results}
\label{sec: findings}

\begin{figure}
    \includegraphics[width=\linewidth]{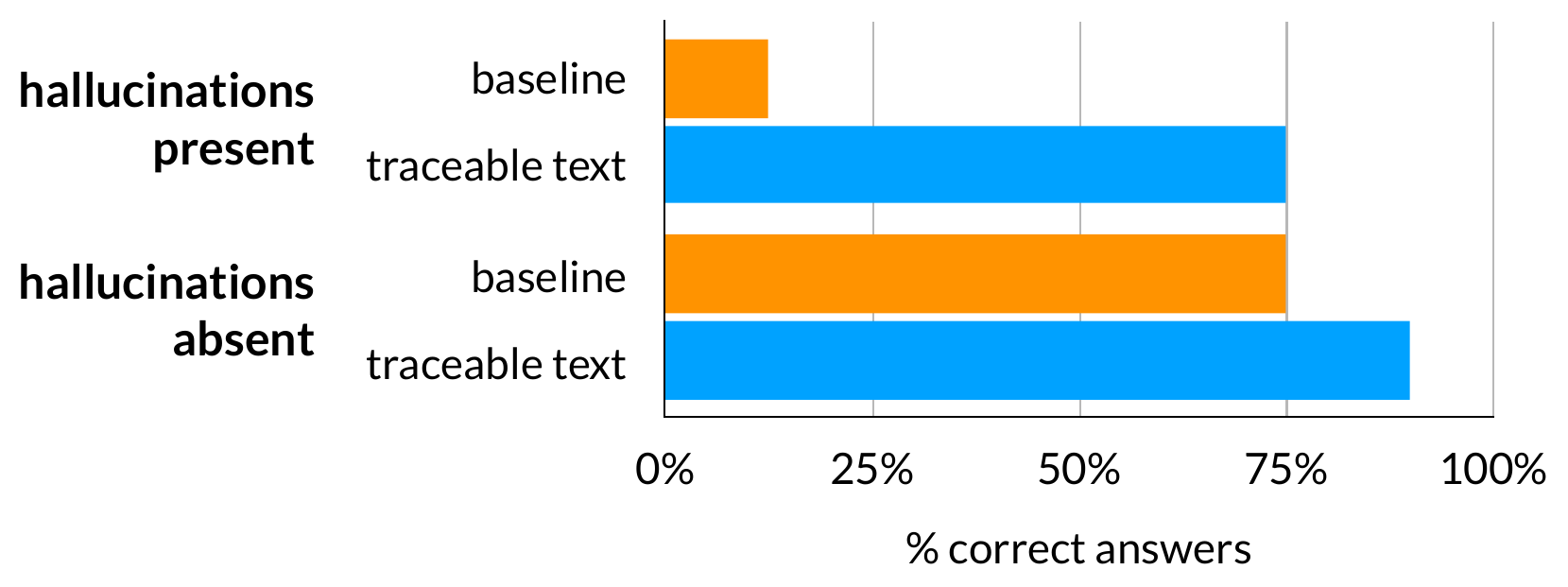}
    \vspace{-4ex}
    \caption{Correctness. \textmd{Readers answered questions correctly more often with traceable text than with the baseline. This effect was pronounced and significant for questions about summaries with hallucinations, and  not statistically significant for questions about expert-validated summaries.}}
    \label{fig:findings-correctness}
\end{figure}

In this section, we answer our three research questions.

\subsection{RQ1. Impact on reading hallucinations}
\label{sec:hallucination-results}

In tasks involving hallucinations, participants answered correctly 70\% of the time with traceable text, and 12.5\% without ($\chi^2 = 27.3$, $p = 1.8 \times 10^{-7}$, Figure \ref{fig:findings-correctness}). They answered questions significantly more quickly, with an average of 1.8 minutes with traceable text ($\sigma = 0.8$) versus 2.9 minutes with the baseline ($\sigma = 1.0$; $F = 45.0$, $p = 3.64 \times 10^{-8}$, see Figure \ref{fig:findings-time}). Despite answering questions markedly more correctly, there was no statistically significant difference in self-reported confidence in one's answer ($W = 721$, $p = .44$) or self-reported difficulty ($W = 919$, $p = .24$).

\subsection{RQ2. Impact on understanding of texts}

Across all tasks (both with hallucinations and verified texts), participants answered questions significantly more correctly when using traceable text ($\chi^2 = 22.2$, $p = 2.4 \times 10^{-6}$). The difference was particularly pronounced for questions involving hallucinations (see last subsection). For the questions involving verified summaries, participants answered  correctly 90\% of the time with traceable text, and 75\% without, though the difference was not statistically significant ($\chi^2 = 3.1$, $p = .08$, see Figure~\ref{fig:findings-correctness}).

\begin{figure}
    \includegraphics[width=\linewidth]{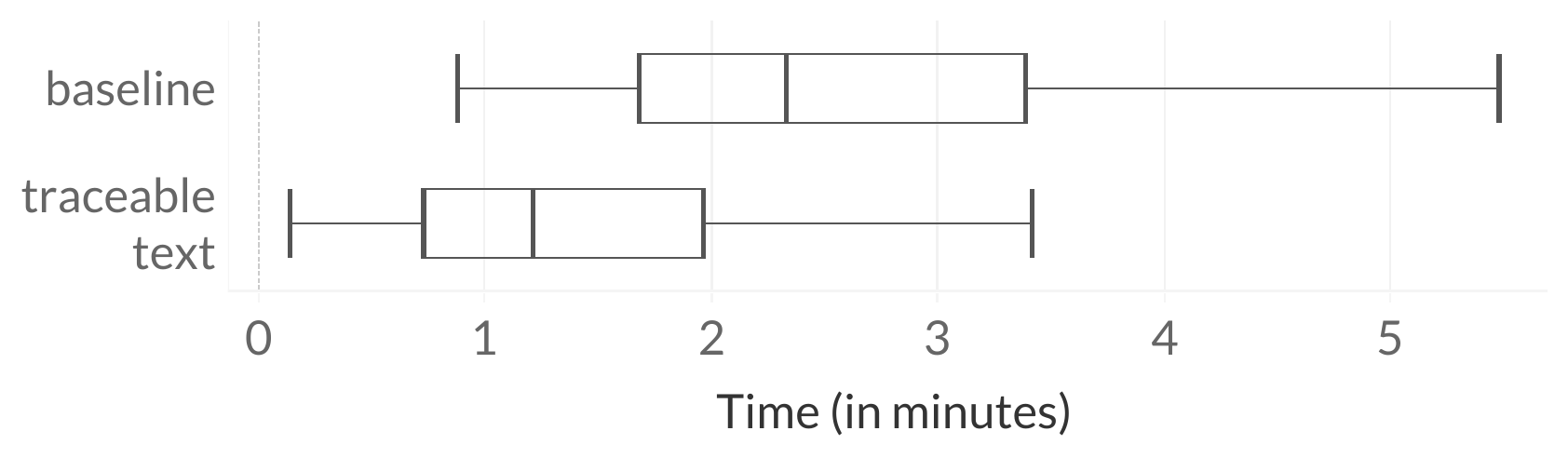}
    \vspace{-4ex}
    \caption{Time to answer. \textmd{Readers answered questions significantly more quickly with traceable text. This effect held both for questions that involved summaries with hallucinations ($F = 45.0$, $p = 3.6 \times 10^{-8}$), and summaries without ($F = 56.1$, $p = 2.0 \times 10^{-9}$).}}
    \label{fig:findings-time}
\end{figure}

Participants answered questions significantly more quickly with traceable text, taking an average of 1.4 minutes  with traceable text ($\sigma = 0.8$) versus 2.5 minutes with the baseline ($\sigma = 1.0$). This difference was statistically significant ($F = 95.4$, $1.2 \times 10^{-16}$, Figure~\ref{fig:findings-time}).

There was no statistically significant difference in the self-reported difficulty of completing tasks with traceable text versus the baseline ($W = 3553$, $p = .22$), or in confidence ($W = 3002$, $p = .49$).

\subsection{RQ3. The experience of using traceable text}
\label{sec:experience}

Participants' impressions of traceable text were positive. In their Likert scale feedback after the unstructured reading task, 10 participants indicated strong agreement and 10 participants indicated agreement to the statement ``I would use this interface to read my progress notes in the future.'' Based on their use in these tasks, participants described
phrase links as ``really cool'' (P15), ``actually really smart'' (P5), and something they ``absolutely love'' (P10).

\paragraph{Value of links}

Participants found several aspects of using phrase links valuable. First, they provided an index into the note. Participants described links as providing ``an entry point to the note'' (P6) and something that made them ``want to read the note a little more'' (P8). The links allowed them to ``zoom in'' on specific phrases within the source document and subsequently ``zoom out'' to the summary, facilitating an understanding of the note's broader context (P2).
Participants also appreciated having links from technical terms in the source text to their simplified forms in the summary ($N=4$).
One participant was even able to find an error in their source note after they read it with the phrase links (P15).

\paragraph{Strategies}

Traceable texts were used in a few ways.
Some participants ($N=6$) read notes with a \emph{summary-first} approach, reading through the summary and using links to refer back to details in the source text. Some of these participants ($N=3$) read through the linked passages exhaustively, exercising every link. In P2's words, this allowed them to ``highlight which aspect of the progress note [each claim in the summary] came from.''

Conversely, participants who indicated that they were more comfortable reading their medical notes ($N=5$) followed a \emph{source-first} approach, using links to clarify information in the note. Two of these participants stated that the phrase links helped them understand otherwise confusing sections of the note (P12, P10).

Some participants ($N=2$) used both summary-to-note and note-to-summary links, largely to check the veracity of the summary.

\paragraph{Limitations}
Participants pointed out limitations to the usefulness of traceable text. Links were seen as less useful when the phrases linked in the summary and note were nearly identical ($N=3$). Furthermore, participants sometimes saw passages in the source text that they wanted to understand, but which were unaddressed in the summary ($N=2$).

\paragraph{Potential extensions}
Participants asked for the ability to link from the summary into multiple documents (such as prior medical notes, P7), and to select sections of the source text for which they could request summaries and links on-demand, if they were not represented in the summary ($N=3$).

\section{Conclusions \& Future Work}

In this paper, we introduce and evaluate traceable text, an interaction primitive that augments an AI-generated summary of source text with phrase-level links between claims in the summary and passages in the source text. A usability study showed traceable text reduced the amount of time it took for readers to answer questions involving inspection of the summary and source text and increased correctness of their answers, particularly when the summary included hallucinations. We also observed readers using traceable text on source texts of personal interest to better understand and index into the source text. Readers used links to read texts in both a summary-first and a source-first fashion. In these ways, traceable text facilitated a deeper engagement with the summary and source document.

\subsection{Limitations}

Our study represents a limited subset of tasks, involving just one kind of source text, and a narrow kind of question. Assessing the utility of traceable text in supporting reading AI-generated summaries more broadly requires evaluation on a broader set of tasks of various domains and levels of complexity.

Although our goal is to deepen reading, our study only covers some aspects of depth---namely, ability to find and reconcile information, and qualitative evidence of use in understanding and indexing into a source text. Additional evaluation is needed to explore other effects like learning and cognitive load, which are affected in nuanced ways by changes to the reading environment.

\subsection{Risks}

As a tool that transforms the information that readers consume, traceable text could have adverse effects on users. This is particularly true in high stakes domains like medicine, which we focus on in this paper. By making source texts easier to read, it could make engagement with them more shallow. By providing an index into those texts, it could disincentivize complete reading of a source text. Additionally, relying on AI models to generate linkages could have adverse effects---if linkages are incorrect, it could lead a reader to believe there is a linkage between an unrelated claim and source passage, and thus overestimate the veracity of a summary claim. These risks require close study before deploying traceable text.

\subsection{Future Work}
Beyond the extensions suggested by participants in our study (Section~\ref{sec:experience}), we see several opportunities for the research community to further explore the potential of traceable text.

\subsubsection{External links} 

Sometimes, external knowledge is necessary to understand a phrase in a summary. For instance, a term may not be defined in a summary, or a source medical note might describe a diagnosis that a patient wants to double-check. Could links support traceability from a source text to external sources that can be used in comprehension and validation? This would likely require different designs, as a reader would not have to collect information from just one short auxiliary text (a summary) but instead potentially multiple, potentially long auxiliary reference texts.

\subsubsection{AI + human workflows for linking}
\label{future:links}

For traceable texts to be used in sensitive domains, further innovations are needed in the AI for generating correct summaries and links, and improved workflows for domain experts to correct them.

\bibliographystyle{ACM-Reference-Format}
\bibliography{references.bib}

\end{document}